\documentclass[twocolumn,showpacs,preprintnumbers,amsmath,amssymb]{revtex4-1}
\usepackage{graphicx}
\usepackage{dcolumn}
\usepackage{bm}
\usepackage{color}
\usepackage{amsmath, amssymb, gensymb}
\usepackage{subfigure}
\usepackage[svgnames*,table,dvipsnames]{xcolor} 
\usepackage{ulem}

\definecolor{RED}{rgb}{1,0,0}

\newcommand{\un}{~\mathrm}
\newcommand{\etal}{\textit{et al.}}

\newlength\epaisLigne
\newcommand{\Ghline}{\noalign{\global\epaisLigne\arrayrulewidth\global\arrayrulewidth 1.5pt}\hline \noalign{\global\arrayrulewidth\epaisLigne}}
\newcolumntype{I}{!{\vrule width 1.5pt}}
\begin{document}
\preprint{APS/123-QED} 

\title{Effect of the porosity on the fracture surface roughness of sintered materials: From anisotropic to isotropic self-affine scaling}
\author{T. Cambonie$^{1}$}\email{cambonie@fast.u-psud.fr,tristan.cambonie@espci.fr}
\author{J. Bares$^{2,3}$}
\author{M.L. Hattali$^{1}$}
\author{D. Bonamy$^{2}$}\email{daniel.bonamy@cea.fr}
\author{V. Lazarus$^1$}
\author{H. Auradou$^1$}
\affiliation{$^1$ Univ Paris-Sud, CNRS, UMR 7608,
Lab FAST, Bat 502, Campus Univ, F-91405, Orsay, France.\\
$^2$ CEA, IRAMIS, SPEC, SPHYNX Laboratory, F-91191 Gif sur Yvette, France\\
$^3$ Present address: Duke University, Durham, North Carolina 27708, USA\\}
%%%%%%%%%%%%%%%%%
\begin{abstract}
To unravel how the microstructure affects the fracture surface roughness in heterogeneous brittle solids like rocks or ceramics, we characterized the roughness statistics of post-mortem fracture surfaces in home-made materials of adjustable microstructure length-scale and porosity, obtained by sintering monodisperse polystyrene beads. Beyond the characteristic size of disorder, the roughness profiles are found to exhibit self-affine scaling features evolving with porosity.
Starting from a null value and increasing the porosity, we quantitatively
modify the self-affine scaling properties from  anisotropic (at low
porosity) to isotropic (for porosity larger than 10\%). 
\end{abstract}

%%%%%%%%%%%%%%%%%%
\maketitle
%%%%%%%%%%%%%%%%

\section{Introduction}

Fractography, i.e. the morphological characterization of \textit{post-mortem} fracture surfaces, is a classical tool used to identify the mechanisms and the damaging processes (fatigue, stress corrosion, cleavage, plastic cavitation, crazing,...) responsible of failure (see \cite{Hull99_book} for recent textbook). 
Since the pioneer work of Mandelbrot \etal \cite{Mandelbrot84_nature}, numerous studies have evidenced the existence of self-affine scaling invariance over a wide range of length-scales (see \cite{Bouchaud97_jpcm,Bonamy11_pr} for reviews). The early measurements in various materials including metallic alloys, ceramics and rocks \cite{Bouchaud90_epl,Maloy92_prl,Schmittbuhl93_grl} reported values for the roughness exponent $\zeta$ close to $0.8$, suggesting the existence of a universal value \cite{Bouchaud90_epl,Maloy92_prl}, independent of the propagation mode and material.
These experimental observations yielded intense theoretical \cite{Hansen91_prl,Bouchaud93_prl,Schmittbuhl95_prl,Larralde95_epl,Ramanathan97_prl,Hansen03_prl,Bonamy06_prl} and numerical \cite{Termonia86_nature,Raisanen98_prb,Nukala06_pre,Nukala10_pre} researches. 

Recently, the picture has been made more complex. Firstly, the roughness exponent has been found to (slightly) depend on the anisotropy of the material microstructure \cite{Morel04_prl} and on the fracture speed \cite{Mallick07_prl}. Secondly, fracture surfaces were shown to exhibit anisotropic scaling features, characterized by two different roughness exponents whether observed along the direction of crack front or along the direction of crack growth \cite{Ponson06_prl,Ponson06_ijf}. Thirdly, fracture surfaces exhibit anomalous scaling \cite{Lopez98_pre,Morel98_pre}; the introduction of an additional global roughness exponent $\zeta_{glob}$ is then necessary to describe the scaling between the global crack width and the specimen size. Fourthly, multiaffinity, disappearing at large scale, was
invoked \cite{Schmittbuhl95_jgr,Santucci07_pre}. Finally, the scale invariance properties were found to depend on the propagation mode: The seminal self-affine feature with $\zeta\simeq 0.8$ is to be linked with quasi-brittle fracture; lower values for $\zeta$, around $0.4-0.5$ \cite{Plouraboue96_pre,Ponson06_prl2,Bouchbinder06_prl,Ponson07_pre} or even logarithmic roughness \cite{Dalmas08_prl} are to be associated with brittle fracture; and a multi-affine regime with a roughness exponent close to $0.5$ is characteristic of ductile fracture \cite{Bouchaud08_epl,Ponson13_ijf,Srivastava14_jmps}. Note that different regimes can be observed on a same fracture surface \cite{Bonamy06_prl,Morel08_pre,Gjerden13_prl}, depending on the scale of observation and its position with respect to the relevant sizes associated with the various dissipation mechanisms (plastic zone size, damage zone size, fracture process zone size). This enables to infer the fracture process zone from a statistical characterization of crack roughness \cite{Ponson06_prb}.  

Most past studies dedicated to the roughness of cracks aimed at characterizing the scale-invariant properties of fracture surfaces. This quest of universality classes, \textit{i.e.} of features independent of the fracturing conditions (loading and material parameters), strayed from the initial metrology purpose of the fractography science. 
Here,  we go back to this primary purpose and seek to characterize how the microstructure affects the fracture surface roughness. 
Rather than using existing materials like ceramics or rocks, we use home-made porous solids obtained by sintering spherical monodisperse grains.
The advantage is that both the grain size and the porosity can be easily adjusted by modifying, respectively, the bead diameter and the sintering parameters, while presenting a structure of cemented grains very comparable to the structures found in natural heterogeneous brittle solids like rocks (sandstones for instance) or concrete, and other artificial heterogeneous brittle solids like pharmaceutical pills or sintered ceramics.
The procedure for preparing the samples is described in detail in Sec.\ref{sec:sintering_protocol}. The samples are finally broken using a technique detailled in  Sec.\ref{sec:fracture} that permits to grow stable Mode I cracks. The fractography of the surfaces (See Sec.\ref{sec:fractography}) is used to determine the propagation mode of the fracture {\it i.e.} inter- or intra-granular propagation. 
The spatial correlation of the roughness of the crack surfaces is, finally, characterized by the structure function of profiles taken along and perpendicular to the direction of crack propagation. The evolution of this function as function of the bead size, the direction of measurements, and the porosity is reported in Sec.\ref{sec:results}. 
Finally, we discuss in Sec.\ref{sec:discussion} the change in the self-affine exponents observed when the porosity is modified.   

\section{Sample preparation and experimental setup}
\label{sec:protocol}
The experiments reported here consist (i) in driving a stable tensile crack with a controlled velocity in a material obtained by sintering polystyrene beads and (ii) subsequently, in analyzing the spatial distribution of crack roughness. The material fabrication, fracture set-up, and topographical recording of the \textit{postmortem} roughness profiles are presented thereafter. 

\subsection{Sintering protocol}
\label{sec:sintering_protocol}
\begin{figure}[htb!]
	\begin{center}
		\subfigure[]{\includegraphics[height=0.3\textwidth]{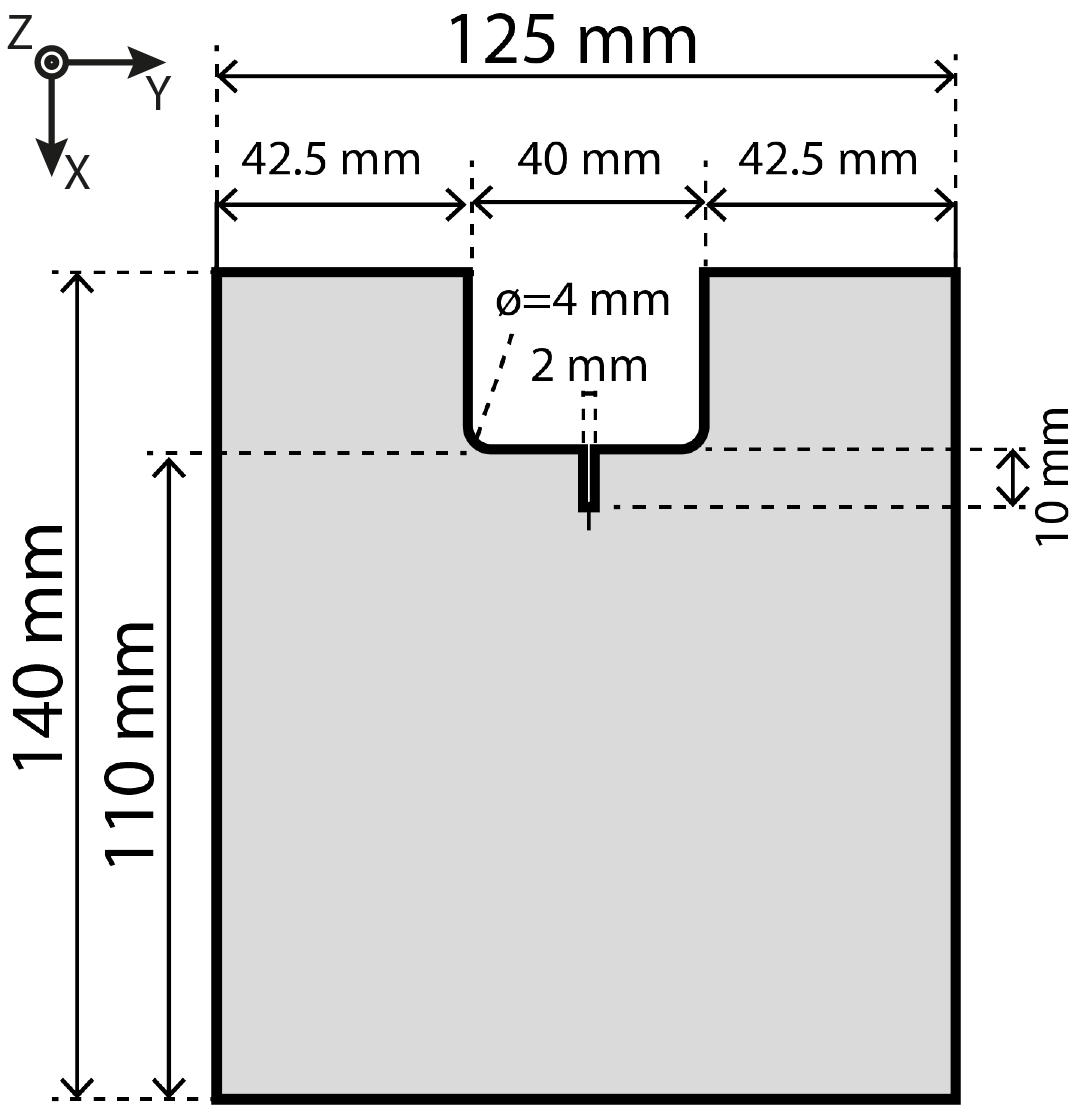}}\\
		\subfigure[]{\includegraphics[height=0.35\textwidth]{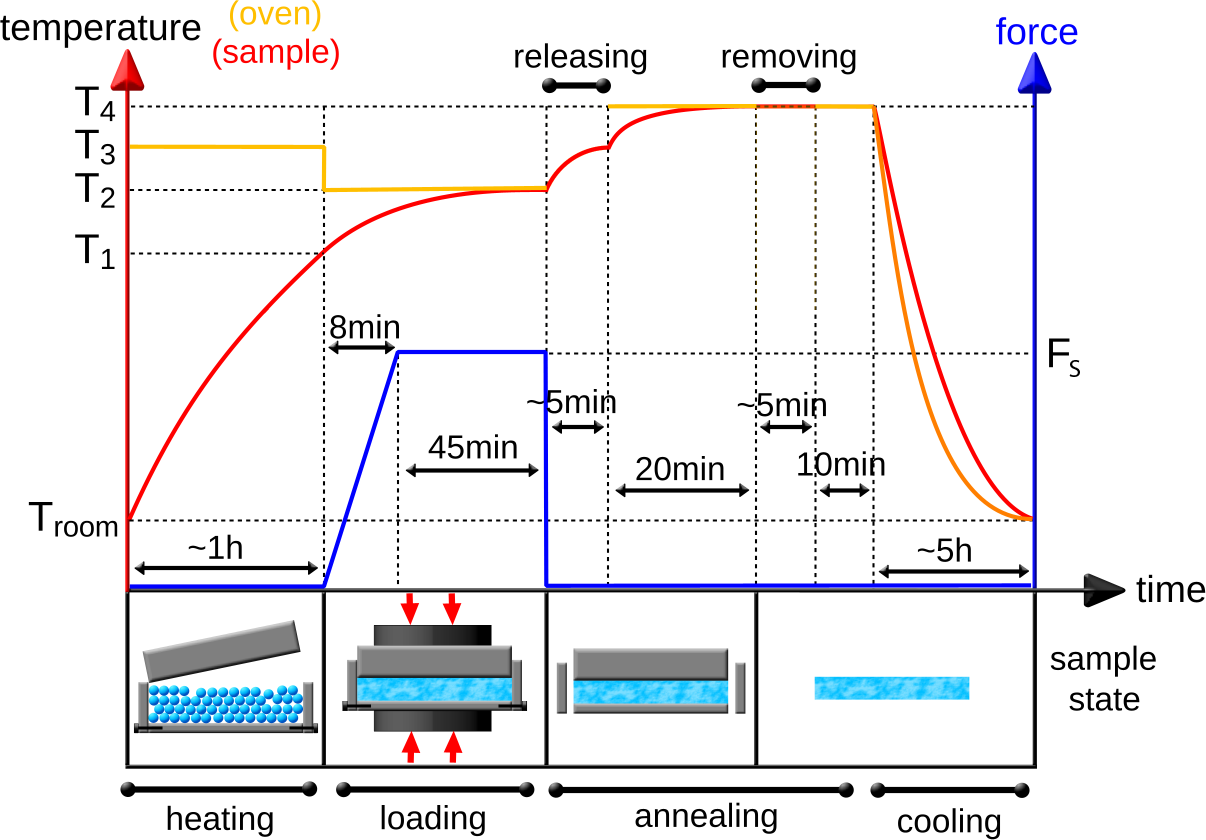}}\\
		\subfigure[]{
\begin{scriptsize}

\begin{tabular}{IcI*{4}{c|}cI}
\Ghline
$d_0$&21 $\mu$m&42 $\mu$m&81 $\mu$m&228 $\mu$m&583 $\mu$m\\\Ghline
$T_1 (\degree C)$&90&90&105&105&105\\\hline
$T_2 (\degree C)$&100&100&115&115&115\\\hline
$T_3 (\degree C)$&105&115&120&120&120\\\hline
$T_4 (\degree C)$&110&120&130&130&130\\\Ghline
\end{tabular}

\end{scriptsize}
}
\end{center}
\caption{a) Geometry and dimensions of the wedge splitting samples. b) (Color online) Evolution sketch of the temperature of the oven, of the sample, of the force applied during the sintering protocol (See \cite{Bares13_phd} for details). c) Temperature value at the different stages of the sintering protocol for the different bead sizes.}
\label{Fig:DispoExp}
\end{figure}

The samples are obtained by sintering monodisperse polystyrene beads (Dynoseeds$^{\copyright}$) of various diameters. 
Four steps are achieved: 
(i)
an home-made mold whose geometry is given in Fig. \ref{Fig:DispoExp}a is filled with beads and heated up
 to $90\%$ of the temperature at glass transition to soften the beads;
 (ii)
 a slowly linearly increasing compressive force is then applied up to the prescribed sintering load $F_{\rm s}$ for nearly one hour to achieve sintering;
 (iii) then, an annealing stage is performed: the mold is unloaded, unscrewed and loosened, keeping the temperature high enough to avoid thermal-shocks;
(iv)
the sample is finally slowly cooled down to ambient temperature. 
The whole sintering protocol is sketched with more details in Fig. \ref{Fig:DispoExp}b, and the relevant temperatures are gathered in Table \ref{Fig:DispoExp}c (see also \cite{Bares13_phd} for an extensive presentation).

The bead diameter $d_0$ was varied between $21$ and $583\ \mu$m. By changing $F_{\rm s}$ between $0.1$ and $8$ kN (corresponding to loading pressure ranging from 6 kPa to 490 kPa), we are able to fabricate specimens of porosities between $\sim 0\%$ and $19\%$ (see Tab. \ref{Tbl:Param}). 
Smaller values of $F_{\rm s}$ leads to unusable friable samples.
The porosity $\Phi$ is measured as the ratio $\Phi=1-\rho/\rho_0$ of the density of the sample $\rho$ and a reference density $\rho_0$ {corresponding to the density of the void free samples obtained with the highest sintering force.
It is found that $\rho_0$ is slightly bead size dependent, which may be explained by a slightly different initial bead material.} 
\begin{table}[htb!]
   \begin{small}
\hspace*{-0.25cm}
\begin{tabular}{Im{2.5cm}I*{5}{c|}cI}
\hline 
\textbf{Experiment n\textsuperscript{o}}&\textbf{1}&\textbf{2}&\textbf{3}&\textbf{4}&\textbf{5}&\textbf{6}\\\Ghline
$\mathrm{d_0}$ ($\mathrm{\mu}$m)&21&21&42&42&42&81\\
\hline
$\mathrm{F_{S}}$ (kN)&4&8&1&8&8&2\\
\hline
$\mathrm{V_{\rm wdg}}$ $(\mathrm{nm.s^{-1})}$&16&16&16&16&16&16\\
\hline
$\Phi$  ($\%$)&1&0&1&0&0&1\\
\Ghline\multicolumn{7}{c}{\vspace*{-0.3cm}}\\\Ghline 
\textbf{Experiment n\textsuperscript{o}}&\textbf{7}&\textbf{8}&\textbf{9}&\textbf{10}&\textbf{11}&\textbf{12}\\
\Ghline
$\mathrm{d_0}$ ($\mathrm{\mu}$m)1&81&81&81&81&228&228\\
\hline
$\mathrm{F_{S}}$ (kN)&4&8&8&8&2&4\\
\hline
$\mathrm{V_{\rm wdg}}$ $(\mathrm{nm.s^{-1})}$&16&16&16&16&16&16\\
\hline
$\Phi$  ($\%$)&1&0&0&0&1&0\\
\Ghline\multicolumn{7}{c}{\vspace*{-0.3cm}}\\\Ghline 
\textbf{Experiment n\textsuperscript{o}}&\textbf{13}&\textbf{14}&\textbf{15}&\textbf{16}&\textbf{17}&\\
\hline
$\mathrm{d_0}$ ($\mathrm{\mu}$m)&228&583&583&583&583&\\
\hline
$\mathrm{F_{\rm s}}$ (kN)&8&8&8&8&8&\\
\hline
$\mathrm{V_{ \rm wdg}}$ $(\mathrm{nm.s^{-1})}$&16&1.6&16&160&1600&\\
\hline
$\Phi$  ($\%$)&0&0&0&0&0&\\
\Ghline\multicolumn{7}{c}{\vspace*{-0.3cm}}\\\Ghline
\textbf{Experiment n\textsuperscript{o}}&\textbf{18}&\textbf{19}&\textbf{20}&\textbf{21}&\textbf{22}&\textbf{23}\\
\hline
$\mathrm{d_0}$ ($\mathrm{\mu}$m)&21&81&42&42&42&42\\
\hline
$\mathrm{F_{S}}$ (kN)&0.1&0.1&0.1&0.2&0.6&0.8\\
\hline
$\mathrm{V_{\rm wdg}}$ $(\mathrm{nm.s^{-1})}$&16&16&16&16&16&16\\
\hline
$\Phi$  ($\%$)&11.65&18.7&15.1&10.4&$\approx$3&$\approx$2\\
\Ghline
\end{tabular}
\end{small}
 \caption{Values of the experimental parameters associated with the different fracture surfaces analyzed here: bead diameter $d_0$ prior to sintering, sintering force $\mathrm{F_{\rm s}}$, wedge velocity $\mathrm{V_{\rm wdg}}$, porosity $\Phi$.}
\label{Tbl:Param}  
\end{table}

\subsection{Fracture wedge-splitting tests}
\label{sec:fracture} 
Specimens (Fig. \ref{Fig:DispoExp}a) are parallelepipeds of dimension $140\times 125 \times W\un{mm}^3$ in the $x\times y\times z$ direction, where the specimen width $W$ depends the material porosity and ranges from $15\un{mm}$ (for the most compacted specimens) to $20\un{mm}$ (for the most porous specimens). A rectangular notch is milled by cutting a $42 \times 30\times W\un{mm}^3$ rectangle from the sample 
at one of the two lateral faces. 
A 10-mm-long 2-mm-thick groove is then introduced in the middle of the rectangular notch using a diamond saw and a seed pre-crack is added at the groove tip by means of a razor blade. This latter operation prevents the propagation of a dynamic fracture and allows to grow slow stable cracks.  
Thereafter, we define the axes $x$ parallel to the direction of crack propagation, $y$ perpendicular to the mean crack plane, and $z$ parallel to the mean crack front.

\begin{figure}[htb!]
	\begin{center}
		\includegraphics[width=0.4\textwidth]{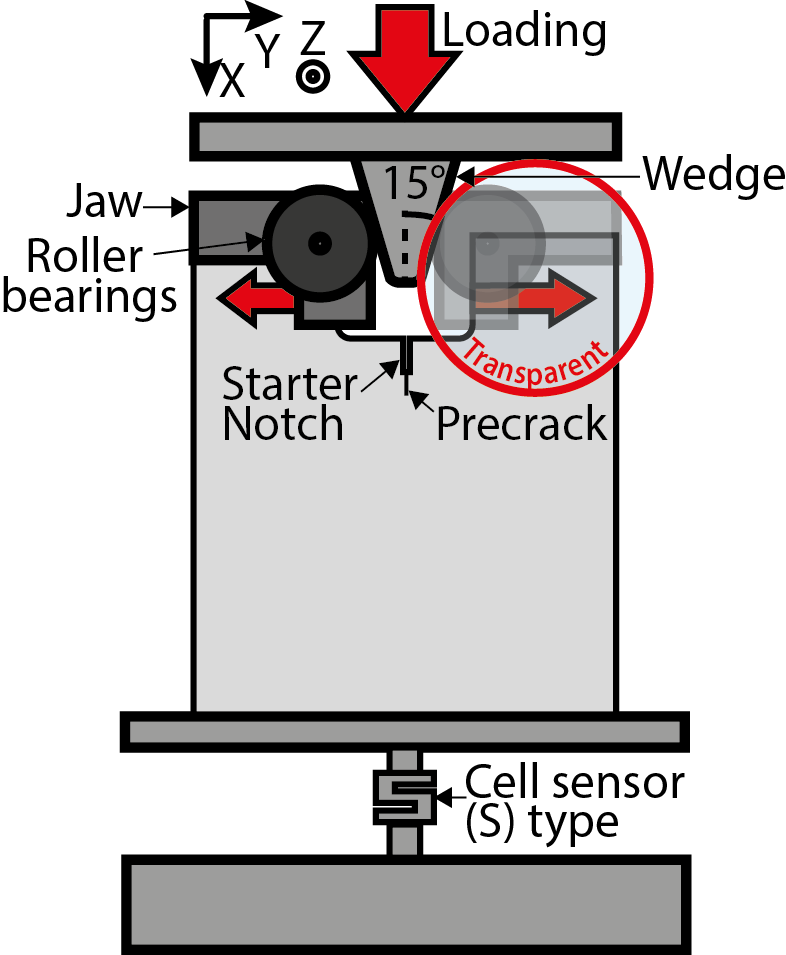}\\a)\\
		\includegraphics[width=0.4\textwidth]{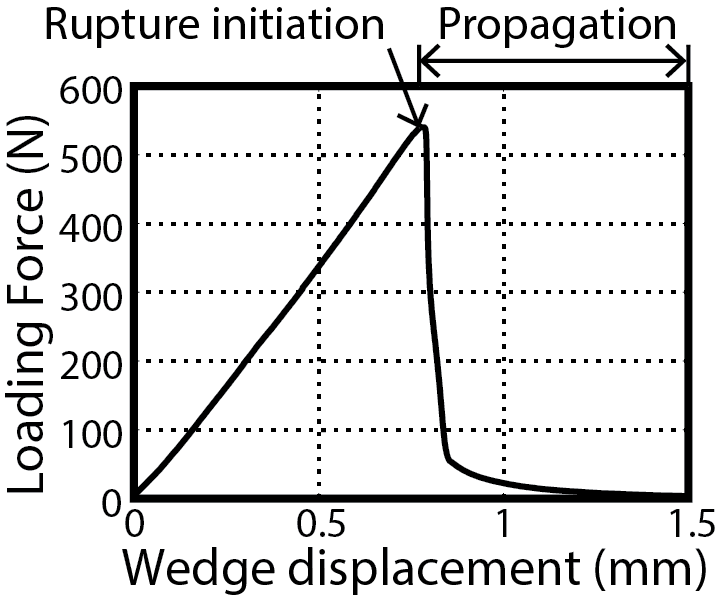}\\b)\\
	\end{center}
\caption{(Color online) a) Experimental setup sketch. b) Typical force $vs.$ displacement curve.}
\label{Fig:DispoExp2}
\end{figure}
 
The sample is then placed between the two jaws of a loading machine (home-made one or commercial electromechanical machine). 
A home-made metallic wedge (semi-angle of $15\degree$) is placed between the jaw and the specimen notch (Fig. \ref{Fig:DispoExp2}a). It has been designed to minimize friction dissipation by means of two rollers and to convert the compression along $x$ into a tension along $y$ \cite{Scheibert10_prl,Guerra12_pnas}. 
The sample is then compressed by lowering the upper jaw at a constant velocity. 
In the experiments reported here, the wedge speed $V_{\rm wdg}$ varied from 1.6 nm.s$^{-1}$ to 1600 nm.s$^{-1}$. During each test, the force $F(t)$ applied by the wedge to the specimen was monitored in real time by a static load cell. As soon as the wedge starts to push on the specimen, $F$ increases. When $F(t)$ reaches the critical loading (Fig. \ref{Fig:DispoExp2}b), the crack starts to propagate. Prior to fracture the load-displacement curve is linear, confirming the elastic behavior of the material.
The crack velocity is observed to first increase (over $\sim 20\un{mm}$), then to stabilize at a roughly constant value, $\sim 100$ times the wedge speed $V_{\rm wdg}$ (over $40$ to $60\un{mm}$), and to, finally, decrease over the last $20\un{mm}$. Our study focus on the interval where the velocity is constant and all measurements are done in that region.

\subsection{Post-mortem analysis of the crack surfaces}
\label{sec:fractography}
Once the samples were broken into two halves, the morphology of the post-mortem crack surfaces were analyzed. Firstly, the surfaces were visualized by means of a numerical microscope Keyence$^{\copyright{}}$. This microscope automatically performs a vertical scan at different heights and reconstructs a well-focused image pixel by pixel over the whole topography range, irrespectively of the level of height variations. 

Topography profiles are also recorded by a Bruker mechanical profilometer. 
Three profiles are scanned along the $x$-direction and 3 others along the $z$-direction. Their lengths are respectively $55\un{mm}$ along $x$, and $14\un{mm}$ along $z$. Each profile has been positioned in the central part of the fracture surface away from the sample borders. They are located between $x=40$ and $95$ mm and at $z=\{-3,0,3\}$ mm for the profiles in the $x$ direction, and between $z=-7$ and 7 mm and at $x= \{-40,60,80\}$ mm for the profiles in the $z$ direction.
The out-of-plane and in-plane resolutions are $\delta y=0.1\un{nm}$ and $\delta x=\delta z=0.8~\mu$m, which corresponds, per profile, to 68750 points in the $x$ direction and 16250 in the $z$ direction.

For $d_0=583\ \mu$m, the fracture surfaces have been measured using another profilometer due to the vertical displacement limitation of the Bruker profilometer \cite{Ponson07_pre}. Its out-of-plane and in-plane resolutions are $\delta y=100\un{nm}$ and $\delta x=\delta z=1~\mu$m. To have a better statistic, ten 75 mm long profiles have been measured in the $x$ direction. 
Yet, in this case, due to the small number of beads in the thickness, no measurements were made in the $z$ direction.

\section{Results}
\label{sec:results}
A general overview of the fracture surfaces obtained by changing the bead size, the porosity and the crack velocity is first given. In most of the cases, the propagation is inter-granular. We focus on this case, investigating the 
effects of the microstructure size and porosity on the fracture surfaces.

\subsection{Fractography} \label{subsection:fracto} 
\begin{figure*}[htbp!]
   \begin{center}
	\hspace*{-0.4cm}\begin{tabular}{p{0.3\textwidth}p{0.3\textwidth}p{0.3\textwidth}}
   	\includegraphics[width=0.3\textwidth]{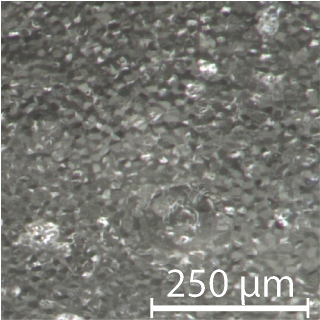}&
    	\includegraphics[width=0.3\textwidth]{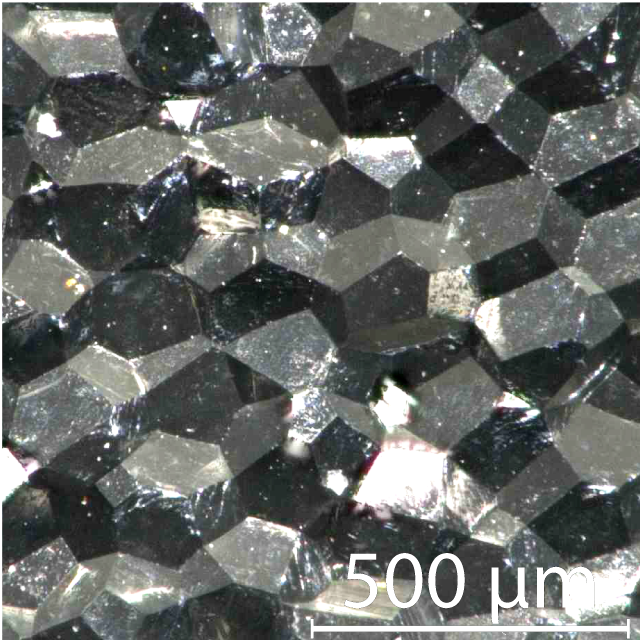}&
    	\includegraphics[width=0.3\textwidth]{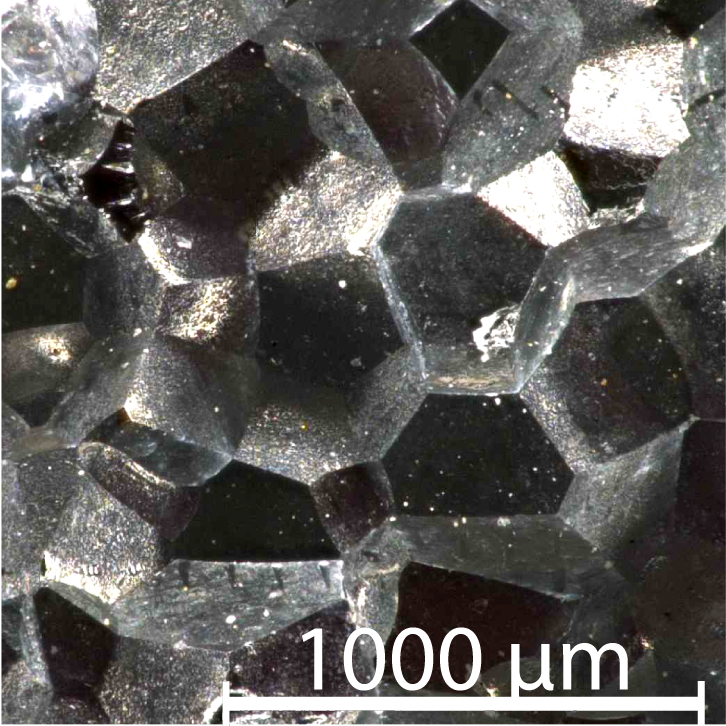}\\
	a) Experiment n$^0$2, $d_0$=21$\mu$m, $\mathrm{V_{\rm wdg}}$=16 $\mathrm{nm.s^{-1}}$, $F_{\rm s}=8$ t, $\Phi=0\%$&
	b) Experiment n$^0$13, $d_0$=228$\mu$m, $\mathrm{V_{\rm wdg}}$=16 $\mathrm{nm.s^{-1}}$, $F_{\rm s}=8$ t, $\Phi=0\%$&
	c) Experiment n$^0$16, $d_0$=583$\mu$m, $\mathrm{V_{\rm wdg}}$=16 $\mathrm{nm.s^{-1}}$, $F_{\rm s}=8$ t, $\Phi=0\%$\\&&\\

    	\includegraphics[width=0.3\textwidth]{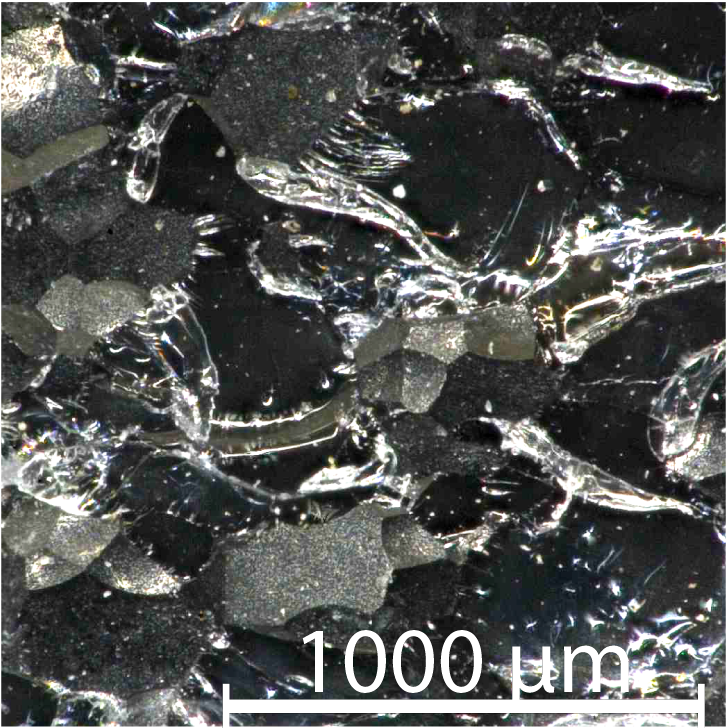}&
  	\includegraphics[width=0.3\textwidth]{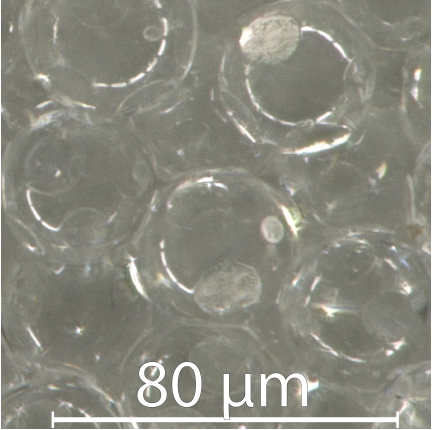}&
    	\includegraphics[width=0.3\textwidth]{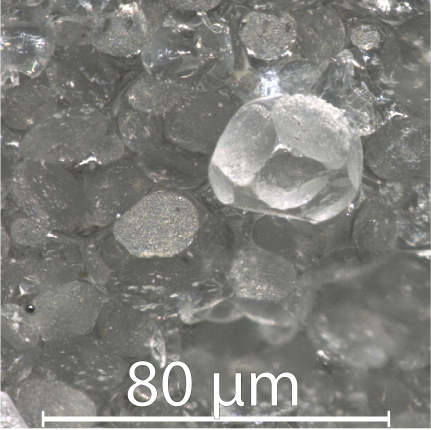}\\
	d) Experiment n$^0$14, $d_0$=583$\mu$m, $\mathrm{V_{\rm wdg}}$=1.6 $\mathrm{nm.s^{-1}}$, $F_{\rm s}=8$ t, $\Phi=0\%$&
	e) Experiment n$^0$21, $d_0$=42$\mu$m, $\mathrm{V_{\rm wdg}}$=16 $\mathrm{nm.s^{-1}}$, $F_{\rm s}$=0.1 t, $\Phi=15\%$&
	f) Experiment n$^0$3, $d_0$=42$\mu$m, $\mathrm{V_{\rm wdg}}$=16 $\mathrm{nm.s^{-1}}$, $F_{\rm s}$=1 t, $\Phi=1.45\%$\\
	\end{tabular}
    \end{center}
\caption{Microscope visualization of wedge-splitted fractured surfaces of the sintered-PS-beads samples.  
The bead diameter $d_0$, the wedge speed $V_{\rm wdg}$, the sintering load $F_{s}$ and the corresponding porosity $\Phi$ are indicated beneath each snapshot.}
\label{Fig:Image_Frittes}
\end{figure*}

Figure \ref{Fig:Image_Frittes} presents the microscope visualizations of typical fracture surfaces, for different bead diameters, porosities and wedge velocities. The crack propagates from left to right.

The snapshots a,b,c correspond to a null porosity and to samples broken at the same wedge velocity, but with different bead diameters. 
The particles surfaces are observed to be completely deformed resulting in sharp edges and corners, in agreement with the zero value of the porosity.
Also, the surfaces are found to be made of smooth facets. Such a faceted morphology is characteristic of a brittle inter-granular fracture, with a crack growing along the grain-grain interfaces. Note that the facets are slightly more elongated along $x$ than along $z$ (visible on Fig. \ref{Fig:Image_Frittes}b). This anisotropy is due to the sintering process. In fact, the compression was applied along $z$, or in a granular packing, the two components $\sigma_{xx}$ and $\sigma_{zz}$ of the stress tensor are classically related \cite{Janssen95_zvdi} via $\sigma_{xx}=K\sigma_{zz}$ where the Janssen constant $K$ is smaller than 1, typically around $0.6$--$0.8$. During sintering, the beads, hence, contract more along $z$ than along $x$.
 The snapshot (d) of Fig. \ref{Fig:Image_Frittes}, corresponding to a smaller wedge speed $V_{\rm wdg}=1.6\un{nm/s}$ presents a qualitatively different morphology: The facets are observed to be blurred with a multitude of small-scale fragments. 
They betray a intra-granular propagation mode, with a crack propagating throughout the grains, without necessary following the grain-grain interfaces. 

Figure \ref{Fig:Image_Frittes}e,f shows microscope visualizations of fracture surfaces of sintered samples realized with lower sintering force $F_{\rm s}$. 
At low $F_{\rm s}$ (Fig. \ref{Fig:Image_Frittes}e), the beads globally keep a spheric shape. The contact areas between adjacent beads is small and are disc-shaped. 
As a result, an important porosity is found in such samples. 
When $F_{\rm s}$ gets larger,  
the contact surfaces grow and intersect, resulting in the apparition of sharp edges of increasing length (Fig. \ref{Fig:Image_Frittes}f), 
  while the  undeformed spherical parts of the particles  shrink, so that the volume of the pores, delimited by these parts, decreases.
  When $F_{\rm s}$ is large enough to close all the pores (Fig. \ref{Fig:Image_Frittes}c), the undeformed parts of the particles disappear and the corners are sharp.

\begin{figure}[htb!]
	\begin{center}
		\begin{tabular}{c}
		\includegraphics[width=0.48\textwidth]{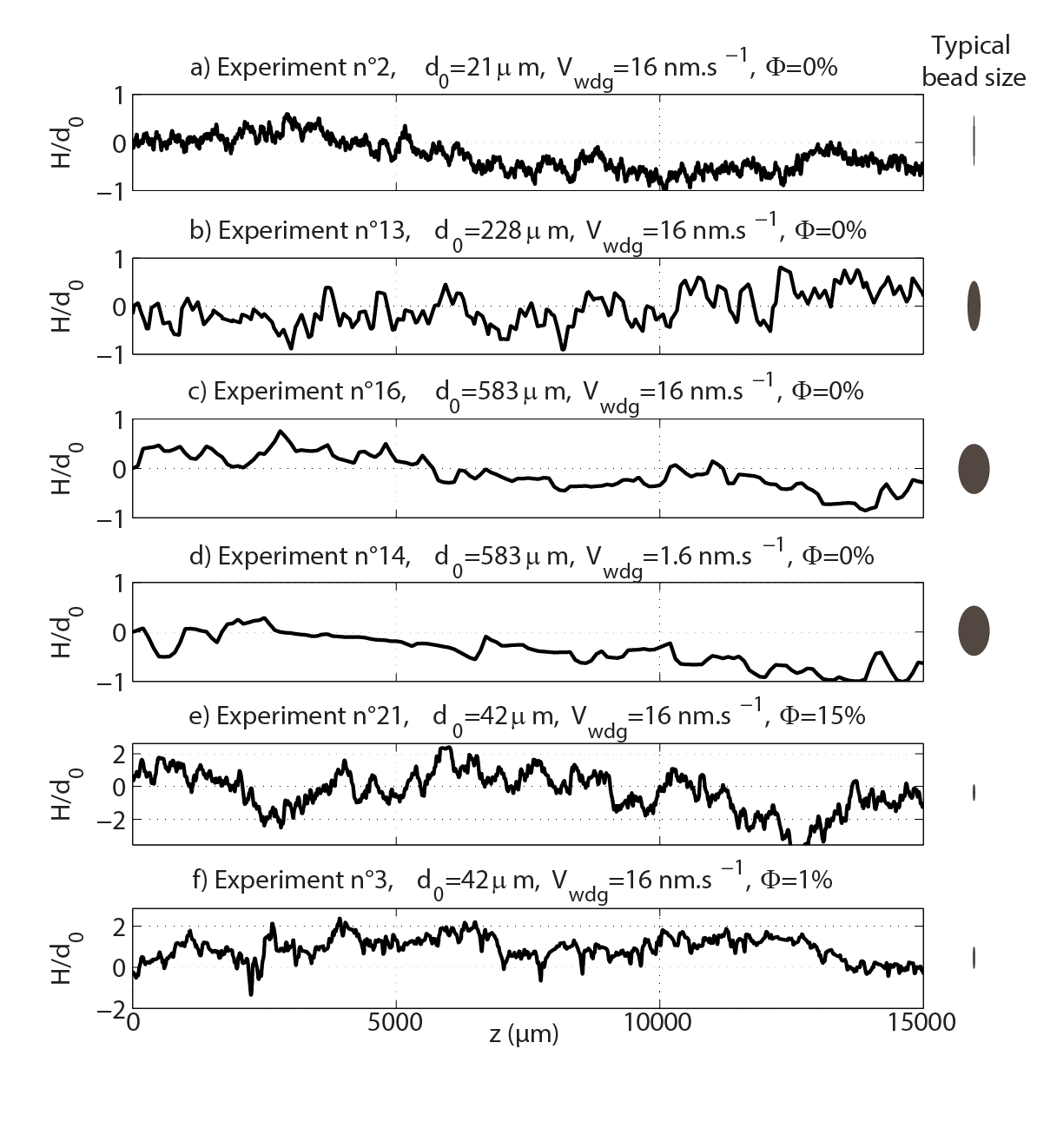}\\
		\end{tabular}
	\end{center}
\caption{Experimental height profiles along the crack propagation direction of the same samples than in Fig.\ref{Fig:Image_Frittes}. 
On the right, the bead shape is given as reference, using the same abscissa and ordinate scales as in the plots. Since these scales are different, the bead appears distorted.}
\label{Fig:ExpProfiles}
\end{figure}

Figure \ref{Fig:ExpProfiles} presents typical topographical profiles  $H(z)$ as measured along the $z$ direction on the fracture surfaces observed in Fig. \ref{Fig:Image_Frittes}. 
In the Fig. \ref{Fig:ExpProfiles}a,b,c associated with inter-granular fracture, the corrugations in the $y$ and $z$ directions are of the same order,
that of the bead size.
 Conversely, the roughness observed for intra-granular mode (Fig. \ref{Fig:ExpProfiles}d) appears much flatter, and exhibits plateaus much longer than the bead diameter.
These plateaus correspond to zones where the crack has cut throughout the grains, without being perturbed by the interfaces. Finally, Fig. \ref{Fig:ExpProfiles}e,f show typical profiles at finite porosity. Note the increase in roughness with amplitudes \textit{larger} than the bead diameter $d_0$. %This is due to the presence of holes in between the sintered grains. 
The shift from inter- to intra-granular fractures occurs for fracture
velocity of 1.6 nm/s. In the present study, we focus on the inter-granular
case, hence on experiments corresponding to $V_{\rm wdg}>1.6$ nm/s. 
Works on the transition between intra- and inter-granular propagation and on the 
intra-granular fracturation are in progress.

\subsection{Effect of the microstructure length-scale}\label{subsection:inter-intra}

We first restrict the analysis to the effect of the microstructure length-scale and, hence, only consider the specimens (1-17 in Tab.~\ref{Tbl:Param}) with very low porosity ($<0.5\%$). 
To quantify the spatial distribution of crack roughness, we computed the structure functions \cite{Bonamy09_jpd} $S_z(\Delta z)$ along $z$, and $S_x(\Delta x)$ along $x$, defined by:
\begin{eqnarray}
S_z(\Delta z) = \langle (H(z+\Delta z)-H(z))^{2}\rangle,\\
S_x(\Delta x) = \langle (H(x+\Delta x)-H(x))^{2}\rangle,
\end{eqnarray}

\noindent where the operator $\langle \rangle$ refers to an average over all possible positions $z$ and $x$, respectively. 

\begin{figure}[htb!]
	\begin{center}
		\begin{tabular}{c}
		\includegraphics[width=0.45\textwidth]{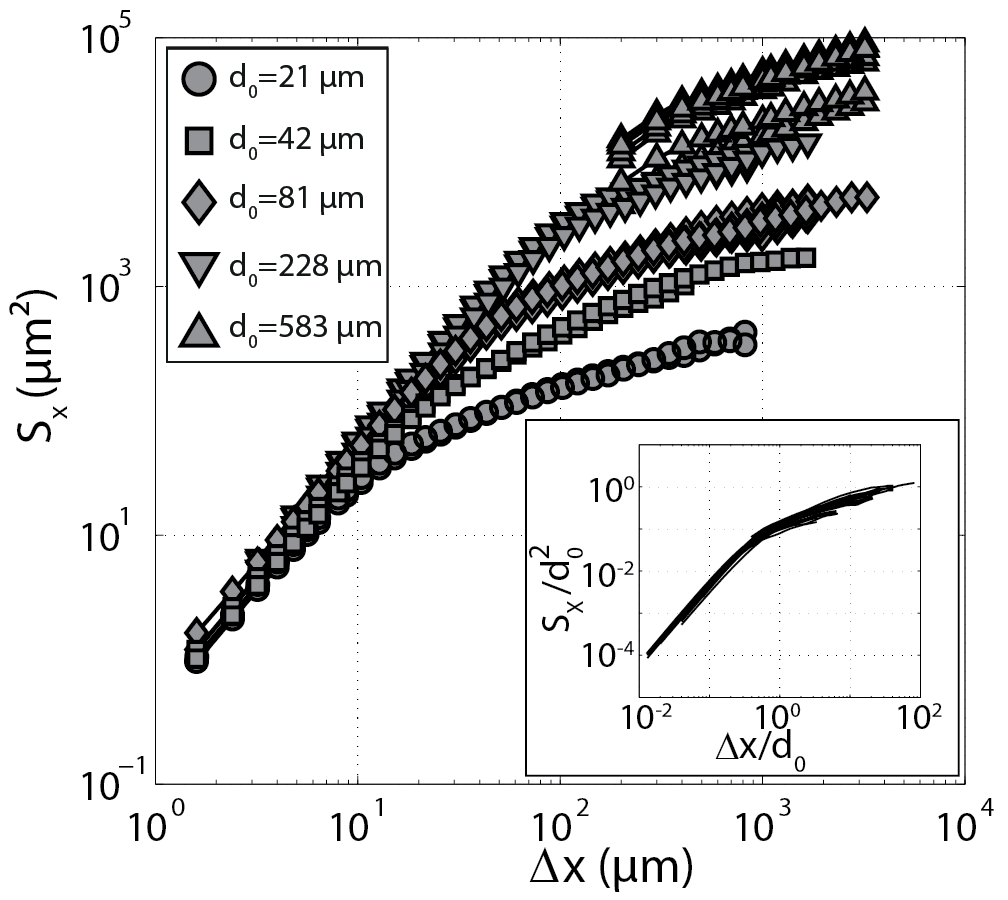}\\
		a)\\ 
		\end{tabular}
	\end{center}
\caption{Structure functions $S_{x}$ of the inter-granular experiments, along the propagation direction at $\Phi=0\%$.  
Each symbol corresponds to a given value of $d_0$.
 Inner window : Same distributions made dimensionless by $d_0^2$.}
\label{Fig:AllProfiles}
\end{figure}

Figure \ref{Fig:AllProfiles} displays in a log-log plot the structure functions $S_x$ measured along $x$ for different bead diameters. Two scaling regimes can be distinguished. At large scale ($\Delta x \gg 10 \mu$m), $S_x$ slowly increases with $\Delta x$, with a prefactor increasing with $d_0$. This betrays the fact that the roughness scales with $d_0$. At small scales ($\Delta x\ll 10 \mu$m), all curves collapse, the roughness is in this regime weakly dependent on $d_0$.

% A SINGLE LENGTH-SCALE IN THE PROBLEM D ... BUT NEED ELLOPT FOR SINTERING REASONS

The diameter $d_0$ appears to be the only length-scale in the problem. Therefore, we made dimensionless the structure functions by making $\Delta x \rightarrow \Delta x/d_0$ and $S_x \rightarrow S_x/d_0^2$ (Inset of Fig. \ref{Fig:AllProfiles}). A good collapse is obtained, confirming that $d_0$ is the length-scale governing the spatial distribution of the roughness.

Nevertheless, the collapse observed in the inset of Fig. \ref{Fig:AllProfiles} is not perfect and a clear scattering remains visible. 
We applied the global minimization technique described in \citep{Cambonie2013} to reduce the dispersion of the data.  
\begin{figure}[htb!]
  \begin{center}
	\begin{tabular}{c}
   	\includegraphics[width=0.45\textwidth]{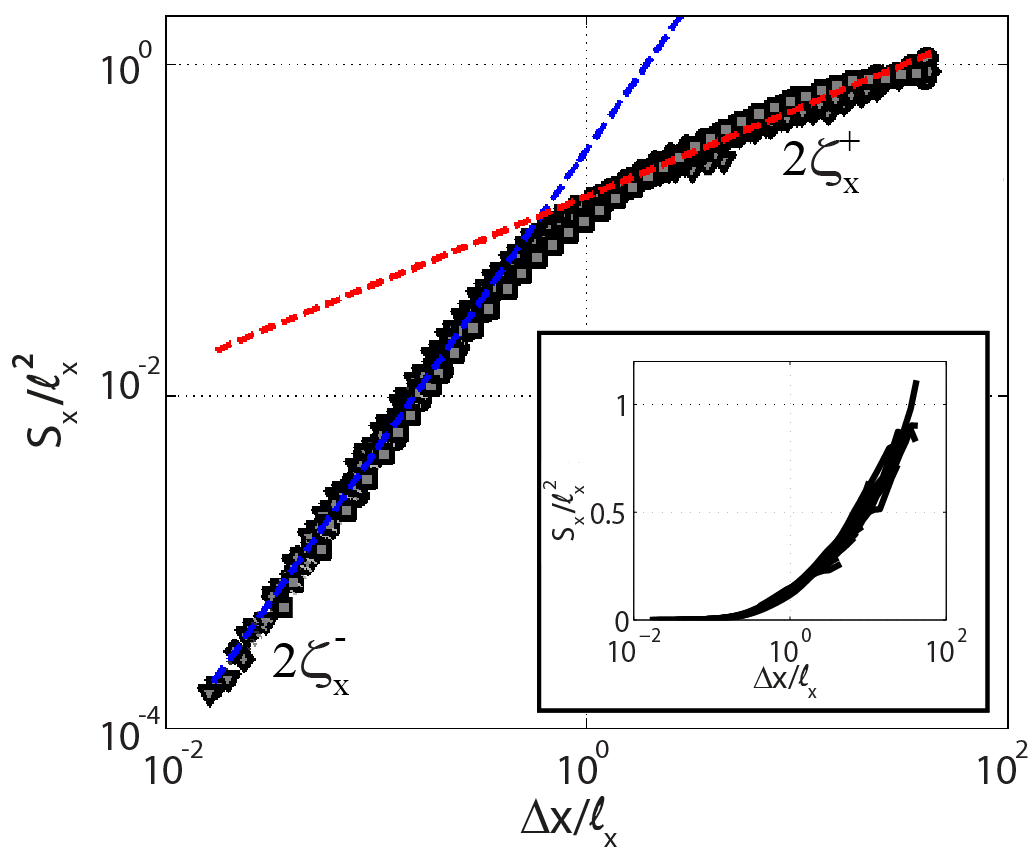}\\
 	a)\\
    	\includegraphics[width=0.45\textwidth]{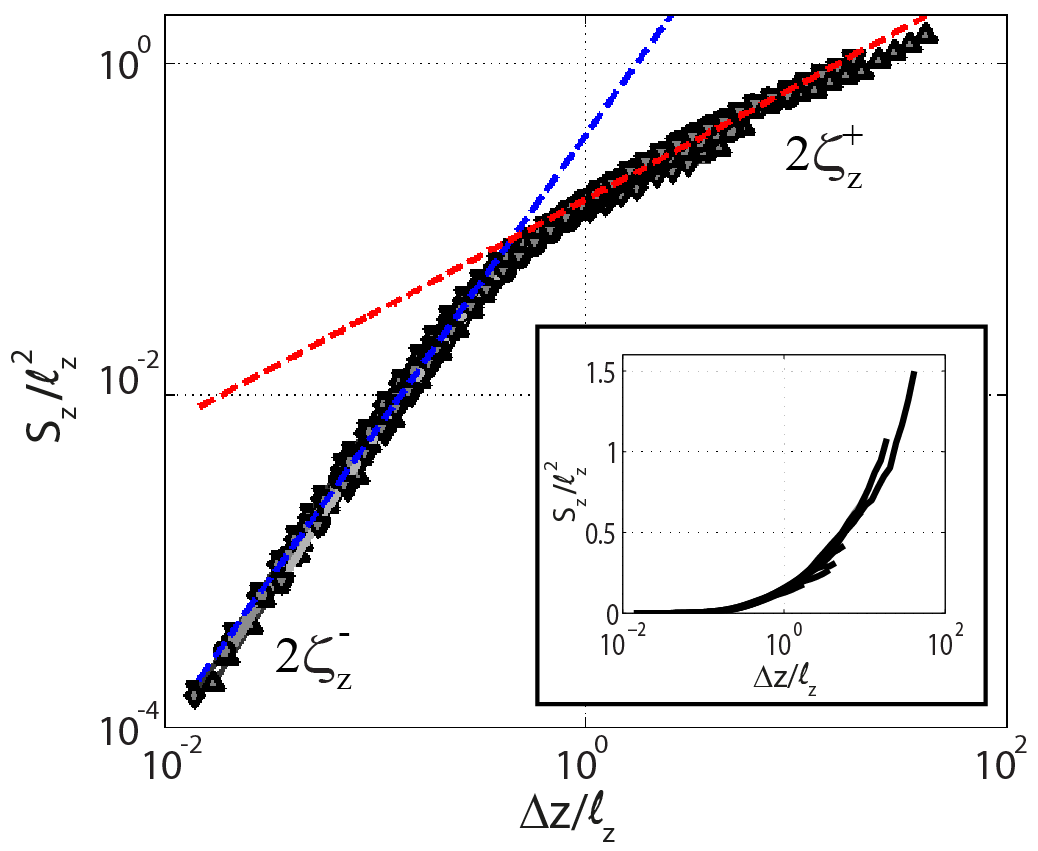}\\
b) \\
	\end{tabular}
    \end{center}
\caption{(Color online) Structures functions normalized by the optimal lengths $\ell_{x}$ and $\ell_{z}$. a) and b) are respectively for the measurements performed along $x$ (propagation direction) and $z$ (front crack). Both axes are logarithmic in the main panels, and semi-logarithmic in inset. The straight lines in the main panels are power-law fits with exponents reported in Tab. \ref{Tbl:exposants}.}
\label{Fig:AdimGProfiles+fit}
\end{figure}
For each curve, an optimal length $\ell$ is obtained.  
The values can be found in Table \ref{Tbl:Param+lopt} for each sample.
Fig. \ref{Fig:AdimGProfiles+fit}a,b shows the structure functions in the $x$ and $z$ directions made normalized by $\ell_x$ and $\ell_z$. As can be seen, the sole parameters $\ell_x$ and $\ell_z$ are sufficient to achieve a very good collapse. We checked that these lengths correspond to the microstructure sizes in both directions (as e.g. observed in fig. \ref{Fig:Image_Frittes}). In other words, the slight differences observed between $\ell_x$, $\ell_z$ and $d_0$ result from the sintering process. 

\begin{table}[htb!]
   \begin{small}
\hspace*{-0.25cm}
\begin{tabular}{Im{2.5cm}I*{5}{c|}cI}
\Ghline 
\textbf{Experiment n\textsuperscript{o}}&\textbf{1}&\textbf{2}&\textbf{3}&\textbf{4}&\textbf{5}&\textbf{6}\\\Ghline
$\mathrm{d_0}$ ($\mathrm{\mu}$m)&21&21&42&42&42&81\\
\hline
$\Phi$ ($\%$)&1&0&1&0&0&1\\
\hline
$\mathrm{\ell_z}$ ($\mathrm{\mu}$m)&19.8&19.6&19.6&39.9&44&79.1\\
\hline
$\mathrm{\ell_x}$ ($\mathrm{\mu}$m)&20.2&21.8&26.4&44.8&44.9&81.7\\
\Ghline\multicolumn{7}{c}{\vspace*{-0.3cm}}\\\Ghline 
\textbf{Experiment n\textsuperscript{o}}&\textbf{7}&\textbf{8}&\textbf{9}&\textbf{10}&\textbf{11}&\textbf{12}\\
\Ghline
$\mathrm{d_0}$ ($\mathrm{\mu}$m)1&81&81&81&81&228&228\\
\hline
$\Phi$ ($\%$)&1&0&0&0&1&0\\
\hline
$\mathrm{\ell_z}$ ($\mathrm{\mu}$m)&70.2&80.3&68.4&76&203.3&187.4\\
\hline
$\mathrm{\ell_x}$($\mathrm{\mu}$m)&74.8&82.9&81.7&83.7&238.7&229.2\\
\Ghline\multicolumn{7}{c}{\vspace*{-0.3cm}}\\\Ghline 
\textbf{Experiment n\textsuperscript{o}}&\textbf{13}&\textbf{14}&\textbf{15}&\textbf{16}&\textbf{17}&\\
\hline
$\mathrm{d_0}$ ($\mathrm{\mu}$m)&228&$\times$&583&583&583&\\
\hline
$\Phi$ ($\%$)&0&$\times$&0&0&0&\\
\hline
$\mathrm{\ell_z}$ ($\mathrm{\mu}$m)&166&$\times$&$\times$&$\times$&$\times$&\\
\hline
$\mathrm{\ell_x}$ ($\mathrm{\mu}$m)&192.9&$\times$&381&486.5&440.8&\\
\Ghline
\end{tabular}
\end{small} 
  \caption{Optimal lengths $\ell_x$ and $\ell_z$ along the $x$ and $z$ directions used to collapse (in Fig. \ref{Fig:AdimGProfiles+fit}) the structure functions.}
\label{Tbl:Param+lopt} 
\end{table}

\begin{table}[htb!]
   \begin{center}
\begin{tabular}{|l|c|c|}
\hline
    Direction & $x$ & $z$ \\\hline
   Small scales : $\zeta^-$& 0.86$\pm$0.04& 0.81$\pm$0.05\\\hline
   Large scales : $\zeta^+$&0.27$\pm$0.02&0.35$\pm$0.01\\\hline
\end{tabular}
\end{center} 
  \caption{Roughness exponents obtained by linear fits performed on the curves of Fig. \ref{Fig:AdimGProfiles+fit}}
\label{Tbl:exposants}  
\end{table}
Moreover, two power-law regimes, characterized by two different exponents, are observed.
To determine these exponents, we performed linear fits of the data displayed in Fig. \ref{Fig:AdimGProfiles+fit}. The roughness exponents $\zeta^{+,-}_x$ and $\zeta^{+,-}_z$, for length scales above and below $d_0$, are defined by the relations $S_x\ \propto\Delta x^{2\zeta_x}$ and $S_z\ \propto\Delta z^{2\zeta_z}$. Table \ref{Tbl:exposants} gives the roughness exponent values along the propagation and front crack directions.
In the small scales regime, the exponents $\zeta^{-}$ are found close to $0.8$ in both directions.
At larger scales, the exponents $\zeta^{+}$ are functions of the directions of measurements, the roughness along the crack being more important than the one in the propagation direction. Although being relatively small, the insets of Fig. \ref{Fig:AdimGProfiles+fit} clearly prove that they are finite and exclude a logarithmic behavior.

\subsection{Effect of the porosity}
\begin{figure}[htb!]
   \begin{center}
\begin{tabular}{c}

    \includegraphics[width=0.5\textwidth]{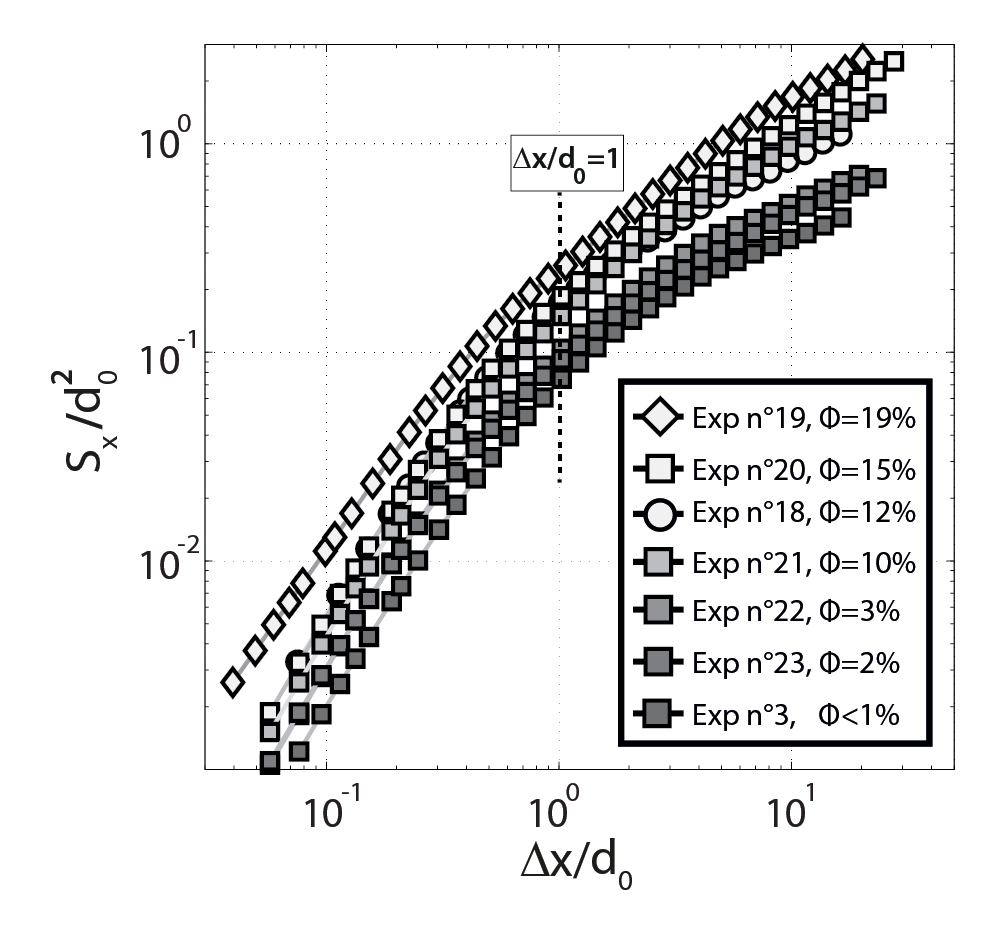}\\
\end{tabular}
    \end{center}\caption{Dimensionless structure functions along the propagation direction for different porosities $\Phi$.}
    \label{Fig:Por}
\end{figure}

We now investigate the influence of the porosity $\Phi$. 
Fig. \ref{Fig:Por} shows the dimensionless structure functions. 
Contrary to the previous case, the data do not collapse. First, the amplitude of the structure functions is found to increase with $\Phi$ at all scales. This is compatible with the observations of rougher profiles as the porosity increases (Fig. \ref{Fig:ExpProfiles}). At small length-scales ($\Delta x/d_0<1$), the exponents are similar for all porosities. A linear regression of the data gives a roughness exponent $\zeta^{-}_x=0.88\pm0.04$, very close to the one otained for $\Phi=0$. 
Yet, for $\Delta x/d_0>1$, the exponents rise with the porosity. 

\begin{figure}[htb!]
   \begin{center}
\begin{tabular}{c}
	\includegraphics[width=0.5\textwidth]{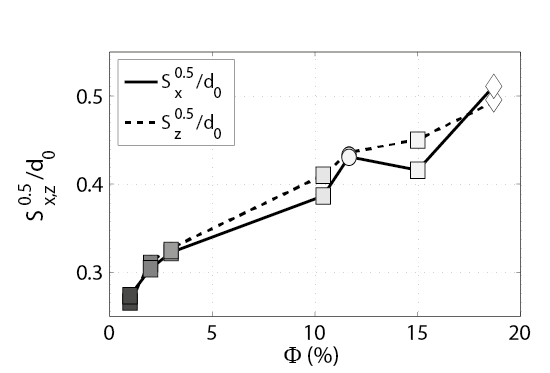}\\a)\\
	\includegraphics[width=0.5\textwidth]{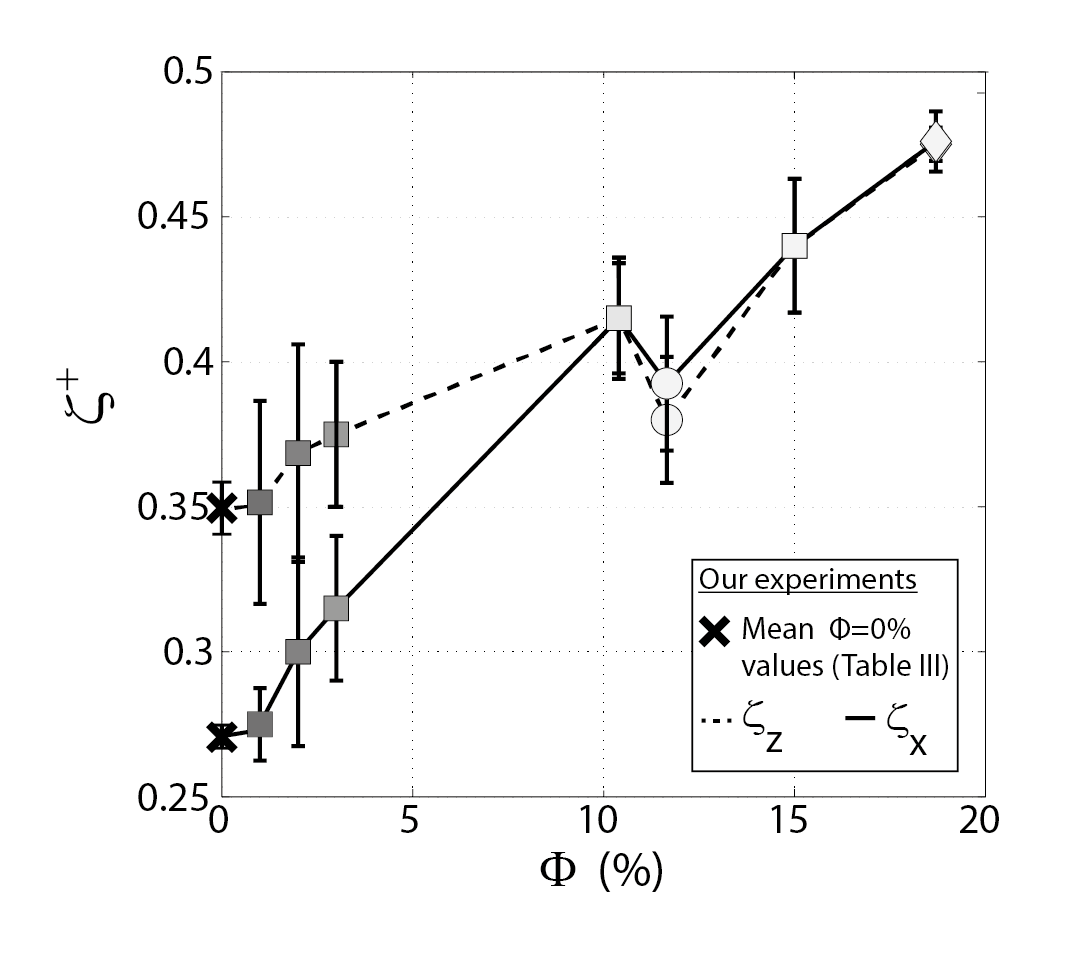}\\b)\\
\end{tabular}
    \end{center}\caption{a) Normalized roughness amplitudes in the $x$ and $z$ directions $\sqrt{S(d_0)}/d_0$ versus porosity $\Phi$, at the position $\Delta x/d_0=1$. b) Large scale roughness exponents $\zeta^{+}$ vs. $\Phi$. Markers match the experiments of Fig. \ref{Fig:Por}. The crosses correspond to the averaged exponent values of Table \ref{Tbl:exposants}.}
    \label{Fig:Por_Quantitative}
\end{figure}

In Fig. \ref{Fig:Por_Quantitative}, we analyze more quantitatively the effect of $\Phi$ on the roughness. The roughness amplitudes, defined as the value of the structure functions at $\Delta x/d_0=1$, are observed to increase linearly with $\Phi$ (Fig. \ref{Fig:Por_Quantitative}a). Fig. \ref{Fig:Por_Quantitative}b shows the evolution of the large scale roughness exponents $\zeta^+_x$ and $\zeta^+_z$ as a function of $\Phi$.
As expected, when $\Phi$ tends toward 0, the values  obtained in the no porosity case (Table. \ref{Tbl:exposants}) are recovered. They increase with $\Phi$, up to values $\zeta^+_z\ \sim \zeta^+_x \sim 0.48$ for $\Phi=20\%$. No significant anisotropy is observed on the roughness amplitude (Fig. \ref{Fig:Por_Quantitative}a). Conversely the scaling anisotropy, i.e. a difference between $\zeta^+_x$ and $\zeta^+_z$, observed for $\Phi=0$ is recovered. However, it decreases as $\Phi$ increases. 

\begin{figure}[htb!]
   \begin{center}
\begin{tabular}{c}
	\includegraphics[width=0.5\textwidth]{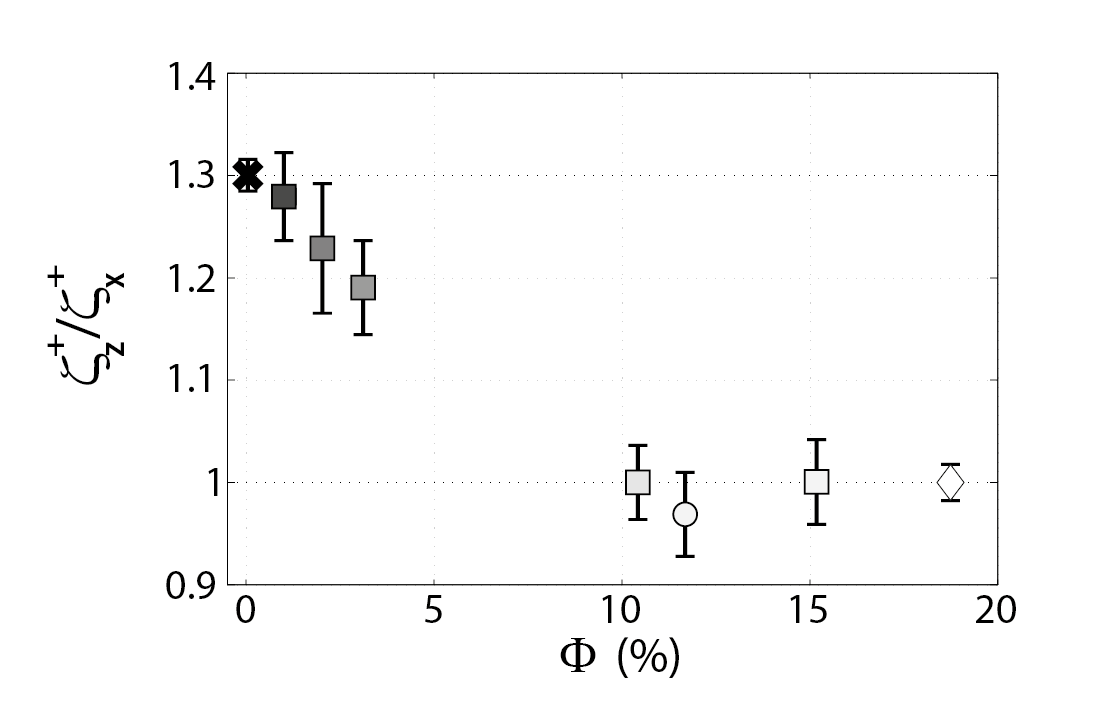}\\
\end{tabular}
    \end{center}\caption{Scaling anisotropy defined as $\zeta^{+}_z/\zeta^{+}_x$ as a function of the porosity $\Phi$.}
    \label{Fig:Anisotropy}
\end{figure}

To quantify the scaling anisotropy evolution, we plot the ratio $\zeta^+_z/\zeta^+_x$ as a function of $\Phi$ in Fig. \ref{Fig:Anisotropy}. It decreases from $\zeta^+_z/\zeta^+_x=1.3\pm0.05$ to 1 as $\Phi$ goes from $0$ to $10 \%$. Above, isotropic scaling is obtained.

\section{Discussion}\label{sec:discussion}

\subsection{Comparison with other experimental results}

The roughness value $\zeta^-$ measured at small scales is observed to be independent of the bead size, the direction measurement, and of the porosity. It is compatible with the $\zeta =0.8$ value reported in the literature for a variety materials \cite{Bouchaud90_epl,Maloy92_prl,Bouchaud97_jpcm}. Here, $\zeta^-$ is observed at a scale smaller than the bead size. This small scale regime is not strictly speaking a self-affine regime: The $\zeta^-$ value can be interpreted as the measured roughness exponent of a piecewise linear profile analyzed at scales smaller than that of its segments (see \cite{Lechenault10_prl} for related analysis). 

Regarding the roughness properties at large scales, it is interesting to compare our results with what has been reported for ceramics obtained by sintering oxide glass beads \cite{Ponson06_prl2}. The roughness amplitude (defined as $\sqrt{S(\delta r=d_0)}$) is found to increase linearly with $\Phi$ as observed in \cite{Ponson06_prl2} for glass ceramics. Conversely, our data converge toward a finite amplitude as $\Phi \rightarrow 0$, while, in glass ceramics, the amplitude was vanishing. The main difference is that, in Ref. \cite{Ponson06_prl2}, decreasing $\Phi$ induces a transition from an inter to intra granular propagation mode, this is not the case here.

The values of the roughness exponents are also, within the errorbars, similar to the values $\zeta \simeq 0.4-0.5$ reported in these glass ceramics \cite{Ponson06_prl2} and in sandstone \cite{Plouraboue96_pre,Ponson07_pre}. Our material is hence more representative of conventional heterogeneous brittle materials such as ceramics and some rocks. As a bonus, the smaller errorbars and the larger range of $\Phi$ explored have evidenced the increase of $\zeta$ with $\Phi$. They have also revealed a scaling anisotropy in absence of any porosity, which decreases as $\Phi$ increases and disappears for $\Phi \geq 10\%$.
 
\subsection{Comparison with theoretical results}
 
Since the early 1990s, a large number of theoretical work has focus on the spatial distribution of fracture surface roughness in brittle heterogeneous materials. They can be classified into two categories: (i) elastic string models that consider the crack front as an elastic line propagating through randomly distributed obstacles \cite{Bouchaud93_prl,Schmittbuhl95_prl,Larralde95_epl,Ramanathan97_prl,Bonamy06_prl,Katzav07_epl}; and (ii) random lattice models that model the material by a network of fuses, springs or beams with randomly distributed breakdowns thresholds \cite{deArcangelis89_prb,Hansen91_prl,Raisanen98_prb,Nukala10_pre}. The main difference between the predictions of these two classes of models is that models of type (i) naturally lead to anisotropic surfaces, where the direction of front propagation plays a specific role, while model of type (ii) lead to isotropic surfaces \cite{Ponson06_prl}.

By essence, elastic line models address situations of nominally brittle fracture, with nonporous materials. Hence, they are the relevant theoretical framework to discuss the observations reported in Sec. \ref{subsection:inter-intra}. In particular, elastic string approaches lead to:
\begin{itemize}
\item[a)] The length scale of the microstructural disorder to be the single relevant length-scale for the structure function; 
\item[b)] Anisotropic surfaces, where the direction of front propagation plays a specific role. 
\end{itemize}
\noindent Both these predictions are in agreement with the experimental observations (Figs. \ref{Fig:AdimGProfiles+fit} and \ref{Fig:Por_Quantitative}). 

\begin{figure}[htb!]
   \begin{center}
\begin{tabular}{c}
    \includegraphics[width=0.45\textwidth]{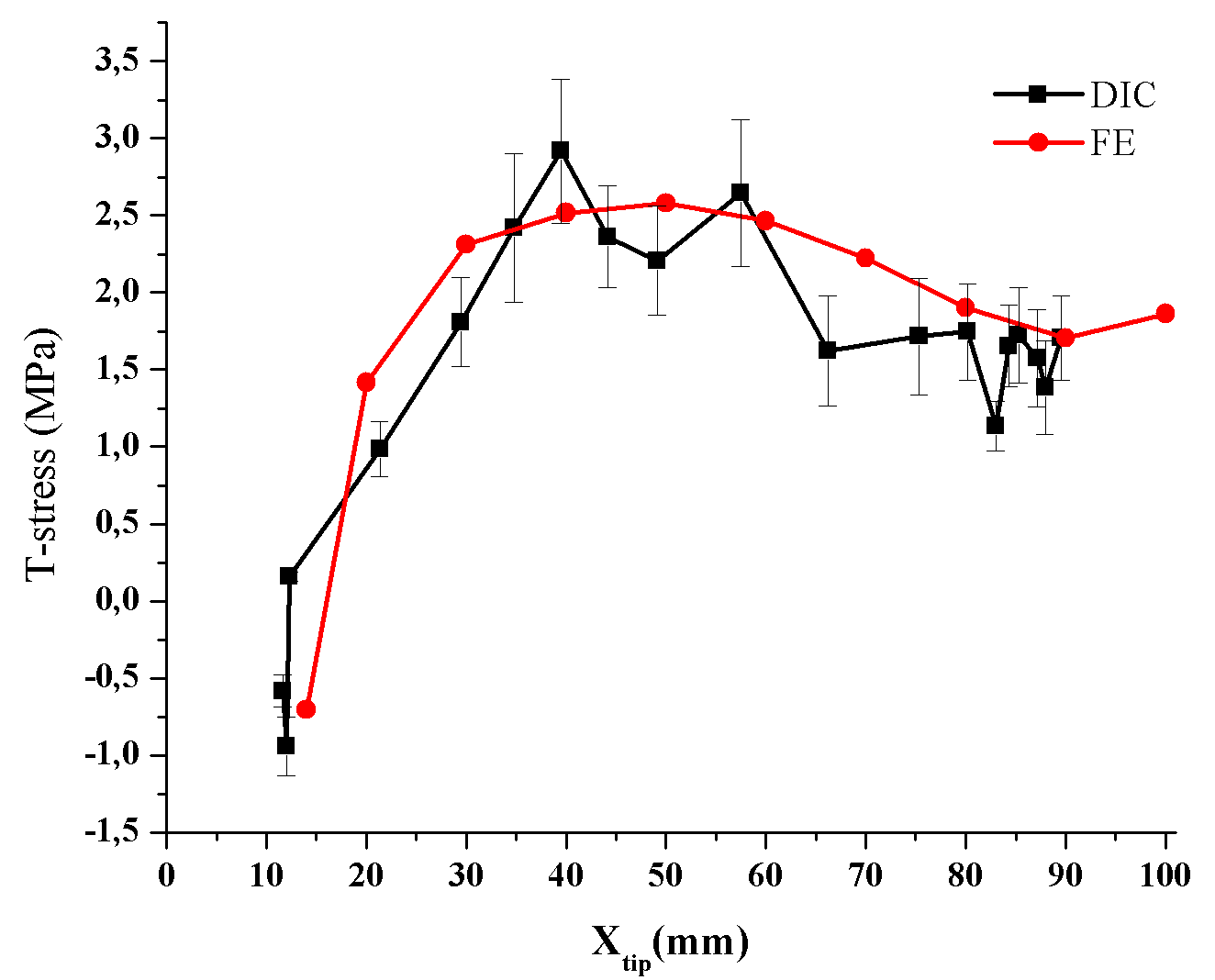}
\end{tabular}
    \end{center}
    \caption{(Color online) $T$-stress as a function of the crack tip position $x_{\rm tip}$ obtained by classical finite element simulations and digital image correlation.}
    \label{Fig:Tstress}
\end{figure}

Conversely, the values of the roughness exponents $\zeta^+_x \approx 0.27 $ and $\zeta^+_z \approx 0.35$ observed here are clearly distinct from the values predicted theoretically. In particular, the most refined models \cite{Ramanathan97_prl,Bonamy06_prl} attempted to derive rigorously the equation of motion of the elastic string from Linear Elastic Fracture Mechanics. They yield either logarithmic scaling \cite{Ramanathan97_prl} or $\zeta_x = 0.50\pm0.05$ and $\zeta_z \approx 0.385\pm0.05$ \cite{Bonamy06_prl} according to the disorder introduced in the equation. The values reported here are clearly different, which indicates that some key ingredients are missing in the models. In particular, no model takes into account the $T$-stress influence. The $T$-stress
here has been determined \cite{ICF13AurHatLaz} both by finite elements simulations and digital image correlation \cite{HilMatRou12}. As shown in Fig. \ref{Fig:Tstress}, the $T$-stress is positive, so that the crack path should be unstable toward any perturbation of its rectilinear path \cite{CotRic80}. In other words, additional correlation are anticipated to result from a positive $T$-stress, and hence a larger value for the roughness exponent measured along $x$. If this scenario is correct, it may partly explain the departure from the logarithmic behavior. 

As mentioned above, in elastic line models, the dimensionless curves $S_x/d^2_0\,\mathrm{vs} \,\, \Delta x/d_0$ and $S_z/d^2_0 \,\, \mathrm{vs} \,\, \Delta z/d_0$ are anticipated to be universal. But, we observed a scaling function of $\Phi$ (See Fig. \ref{Fig:Por}). This indicates that elastic line approaches stop being relevant as the porosity becomes finite. This is thought to be due to the presence of microcracks forming at the pores, and making the fracture propagation mode shifting gradually from nominally to quasi-brittle. The anisotropy decrease with $\Phi$ is consistent with this scenario: As microcracking develops, the fracture surface stop being the trace left by a single propagating line in an otherwise intact material, but result from the coalescence of multiple microcracks instead. Random lattice models include such processes and, may be the relevant framework to describe this quasi-brittle regime. In this context, it is interesting to note that the roughness exponents $\zeta_x^+=\zeta_z^+ \approx 0.45$ measured here on isotropic fracture surfaces, i.e. above $\Phi=10\%$, are consistent with the values $\zeta_{loc}=0.42$, $\zeta_{loc}=0.5$ and $\zeta_{loc}=0.48$ observed respectively in 3D random fuse \cite{Raisanen98_prb}, spring \cite{Parisi00_epl} or beam \cite{Nukala10_pre} networks. 

In the above scenario, porosity provides a tunable parameter to go gradually from nominally brittle fracture, where the disorder effect is to distort the front propagation, to quasi-brittle fracture where the fracture surfaces emerge from a percolating path throughout the microcrack cloud. At present, the theoretical descriptions of these two situations belongs to two distinct realms : Elastic line models for the former and random lattice models for the latter. Unifying these two frameworks represent an important challenge for future investigations (see \cite{Gjerden13_prl} for a recent thrust in this direction). 

\section{Conclusion}

In this paper, we have used home-made cemented grains materials to investigate the influence of the microstructure size and porosity on the fracture surfaces. The following observations have been made: 
\begin{itemize}
\item The roughness displays scale invariant morphological features that depends on both the grain size and porosity.
\item At zero porosity, the structure functions measured along a direction and normalized by the grain size collapse onto a single master curve.
\item Scaling anisotropy is well pronounced for $\Phi=0$, but decreases with $\Phi$ and vanishes for $\Phi>10\%$. 
\item The values of the roughness exponents increase with $\Phi$, from $\zeta^+_x\sim 0.27$ and $\zeta^+_z\sim 0.35$ at $\Phi=0\%$ to $\zeta^+_x\sim \zeta^+_z\sim 0.48$ at $\Phi=20\%$. 
\end{itemize}

These observations are compatible with previous studies considering other sintered glass beads materials \cite{Ponson06_prl2} or sandstone \cite{Plouraboue96_pre,Ponson07_pre}. But in these previous studies, the precision was not sufficient to detect any clear variation with $\Phi$. 
However, 
the values of the roughness exponents $\zeta^+_x \approx 0.27 $ and $\zeta^+_z \approx 0.35$ observed, at low porosity, are clearly distinct from the values predicted theoretically : additional investigations are necessary to take into account some missing ingredients like the $T$-stress. 
Conversely, for $\Phi>10\%$, the roughness exponents $\zeta_x^+=\zeta_z^+ \approx 0.45$ are consistent with the values $\zeta_{loc}$ between $0.4$ and $0.5$, obtained with random networks \cite{Raisanen98_prb,Parisi00_epl,Nukala10_pre}. Modulating $\Phi$ has permitted, thus, to modify quantitatively the value of the roughness exponents. Note that, in conventional interface growth problems, the roughness exponent value is characteristics of a universality class. Their continuous evolution  with $\Phi$ observed herein may pose a rather severe test for current and future competing models of heterogeneous fracture. 
The $T$-stress component acting on the fracture is controlled by the macroscopic shape of the porous samples; future works will consider the effect of the $T$-stress on the roughness.

\begin{acknowledgements}
 We acknowledge support by the ANR Mephystar (ANR-09-SYSC-006-01). 
 We acknowledge P. Viel for the use of the Bruker profilometer and T. Bernard for technical support. 
 We thank R. Pidoux and L. Auffray and T. Bernard for realizing the samples.
\end{acknowledgements} 
\bibliographystyle{unsrt}
 
\bibliography{bibfracture}

\end{document}